\documentclass[twocolumn,secnumarabic,amssymb,nobibnotes,aps,superscriptaddress,pra]{revtex4-2}
 
\usepackage[dvips]{graphicx}
\usepackage{bm}
\usepackage{latexsym} 
\usepackage{amsmath}
\usepackage{mathtools}
\usepackage{empheq}
\usepackage{cases} 
\usepackage{ulem}
\usepackage{comment}

\usepackage{amssymb}
\usepackage{color}

\newcommand{\bmH}{{\bm H}}

\newcommand{\bmm}{{\bm m}}
\newcommand{\bmn}{{\bm n}}

\newcommand{\bmS}{{\bm S}}
\newcommand{\bmsig}{{\bm \sigma}}

\newcommand{\bra}{\langle}
\newcommand{\ket}{\rangle}
\newcommand{\kB}{k_{\rm B}}

\makeatletter
\renewcommand*{\p@subsection}{}
\renewcommand*{\p@subsubsection}{}
\makeatother

\begin{document}

\title{
  Spin Seebeck effect in two-sublattice ferrimagnets in the vicinity of $T_{\rm C}$ 
}

\author{Hayato Fukushima} 
\affiliation{Department of Physics, Okayama University, Okayama 700-8530, Japan}

\author{Masanori Ichioka}
\affiliation{Research Institute for Interdisciplinary Science, Okayama University, Okayama 700-8530, Japan}
\affiliation{Department of Physics, Okayama University, Okayama 700-8530, Japan}
\author{Hiroto Adachi}
\affiliation{Research Institute for Interdisciplinary Science, Okayama University, Okayama 700-8530, Japan}
\affiliation{Department of Physics, Okayama University, Okayama 700-8530, Japan}
\date{\today}

\begin{abstract}
  Spin Seebeck effect refers to the magnonic thermal spin injection from a magnet into the adjacent heavy metal. A ferrimagnetic insulator yttrium iron garnet (YIG) is the material most studied for the spin Seebeck effect. Here, to account for a convex downward temperature dependence of the spin Seebeck effect observed in YIG/Pt system near the Curie temperature $T_{\rm C}$, we develop Ginzburg-Landau theory of the spin Seebeck effect in two-sublattice ferrimagnets. We find that only when we take into account the ``N\'{e}el coupling'', i.e., interfacial exchange coupling between the N\'{e}el order parameter of YIG and spin accumulation of Pt, the convex downward temperature dependence is explained. The present result sheds light on the importance of the N\'{e}el coupling in ferrimagnetic spintronics. 

\end{abstract}



\maketitle

\section{Introduction \label{Sec:I}}

The spin Seebeck effect (SSE)~\cite{Uchida08,Uchida10a} constitutes an important part of spin caloritronics~\cite{Bauer12}. The SSE has now been established as a simple and versatile means for spin injection from magnets~\cite{Uchida14}. Moreover, the SSE has been recognized as a new probe of unconventional magnetic orders~\cite{Hirobe17,Hirobe19,YChen21,Nakagawa25,Matsushita25,Kato25}. In the history of the SSE research, a ferrimagnetic insulator yttrium iron garnet (Y$_3$Fe$_5$O$_{12}$, YIG) has played an outstanding role. Indeed, due to its very low damping~\cite{Serga10}, YIG has been a prototypical platform for the SSE as seen in a number of publications for longitudinal SSE~\cite{Uchida10b}, acoustic spin pumping~\cite{Uchida11}, spin thermoelectric coating~\cite{Kirihara12}, length scale in the SSE~\cite{Rezende14,Kehlberger15}, SSE in the nonlocal configuration~\cite{Cornelissen15}, magnon-polarons in the SSE~\cite{Kikkawa16}, enhanced SSE by antiferromagnetic insertion~\cite{Lin16}, time-resolved SSE~\cite{Schreier16,Kimling17}, spin colossal magnetoresistance~\cite{Qiu18}, and so on. 

Although YIG is a ferrimagnet possessing many magnon branches~\cite{Harris63,Cherepanov93,Princep17,Shamoto18}, most of the experiments listed above are explained using the picture of thermal spin pumping caused by simple ferromagnetic magnons~\cite{Xiao10,Adachi10,Adachi11,Adachi13}. This is because these experiments are done below room temperature, where thermally excited states are dominated by the acoustic (ferromagnetic) magnons. However, as demonstrated by neutron scattering experiment~\cite{Plant77} and atomistic spin dynamics simulation~\cite{Barker16}, upon approaching the Curie temperature $T_{\rm C}$, the optical (antiferromagnetic) magnon mode obtains an energy comparable to thermal energy, such that this mode cannot be neglected as the antiferromagnetic mode carries spins with opposite helicity to the ferromagnetic mode. 

Experimentally, Uchida {\it et al.} measured the SSE in a YIG/Pt system at high temperatures and observed that the SSE signal has a convex {\it downward} temperature dependence near $T_{\rm C}$~\cite{Uchida14b}. Theoretically, Barker and Bauer conducted an atomistic thermal spin dynamics simulation for YIG and found that the SSE signal shows a convex {\it upward} temperature dependence near $T_{\rm C}$~\cite{Barker16}. Subsequently, Ginzburg-Landau (GL) theory of the SSE in a simple ferromagnet was developed, and it was concluded that the SSE signal exhibits a convex {\it upward} temperature dependence near $T_{\rm C}$~\cite{Adachi18}, whose result is consistent with the atomistic numerical simulation~\cite{Barker16}. Therefore, it is highly desirable to resolve this long-standing conflict between experiment and theory.

In this paper, we develop GL theory of the SSE in two-sublattice ferrimagnets near $T_{\rm C}$. The GL theory is an effective model for the underlying spin Hamiltonian, and is valid in the vicinity of a phase transition~\cite{Bellac-text}. Equilibrium GL model has been successfully applied to the static critical phenomena~\cite{Ma-text}. The time-dependent Ginzburg-Landau (TDGL) equation is a dynamical generalization of the equilibrium GL equation, and it has been applied to the ferromagnetic SSE~\cite{Adachi18} as well as to the antiferromagnetic SSE~\cite{Yamamoto19,Yamamoto22}. In this paper, we first construct GL model of two-sublattice ferrimagnets, which contains two order parameters of magnetization vector and N\'{e}el vector. Next, we generalize it to the TDGL theory and describe the SSE in a ferrimagnet/heavy metal bilayer.

In our GL approach, the important parameters are the two types of exchange coupling at the ferrimagnet/heavy metal interface, which we term either ``magnetic coupling'' or ``N\'{e}el coupling''. The magnetic coupling is the interaction between the magnetization vector and spin accumulation at the ferrimagnet/heavy metal interface, and this coupling describes the usual interfacial exchange coupling~\cite{Adachi18}. On the other hand, the N\'{e}el coupling is the interaction between the N\'{e}el vector and spin accumulation, which is considered to survive only for a spin-uncompensated interface~\cite{Tang24}. Inspired by the phenomenological approach to the exchange-bias effect, the N\'{e}el coupling was proposed in the early stage of antiferromagnetic spintronics~\cite{Takei15} [see Eq. (1) therein]. More recently, the N\'{e}el coupling was used in Ref.~\cite{Yamamoto22} to explain the strange sign change of the antiferromagnetic SSE across the spin-flop transition~\cite{Li20}. Below, we show that this N\'{e}el coupling also provides a key to understand the ferrimagnetic SSE near $T_{\rm C}$. Namely, we demonstrate that only when the N\'{e}el coupling is dominant, the SSE is dominated by the N\'{e}el order parameter fluctuations, and the convex downward temperature dependence of the SSE observed in YIG/Pt system~\cite{Uchida14b} can be explained.

This paper is organized as follows. In Sec.~\ref{sec:II}, we develop a formalism which allows us to describe the SSE in two-sublattice ferrimagnets near $T_{\rm C}$, and also show our numerical results for temperature dependence of magnetization and N\'{e}el order parameter, acoustic (ferromagnetic) mode and optical (antiferromagnetic) mode. In Sec.~\ref{sec:III}, we calculate temperature dependence of the SSE signal near $T_{\rm C}$. Finally in Sec.~\ref{sec:IV}, we discuss and summarize our results.

\section{Formulation \label{sec:II}  } 
\subsection{Model}

\begin{figure}[t] 
  \begin{center}
    \includegraphics[width=8.5cm]{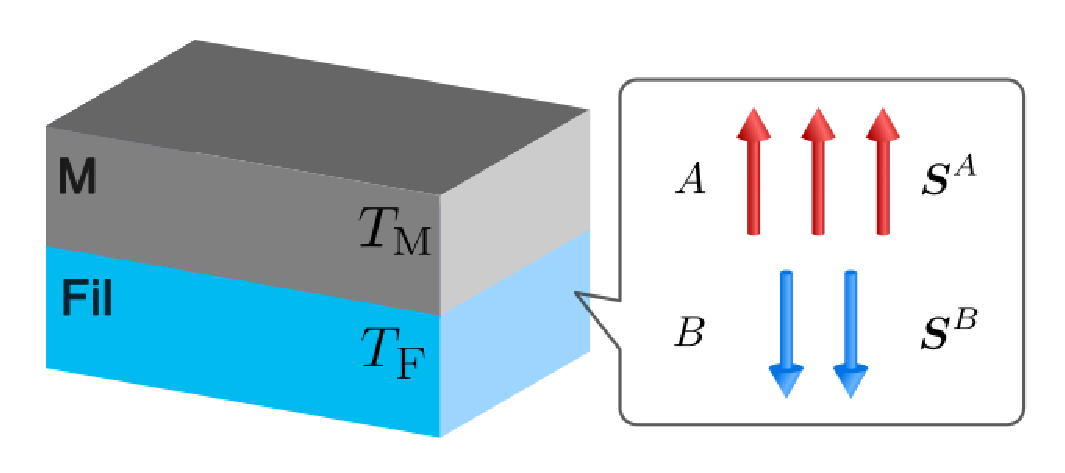}
  \end{center}
  \caption{
    Schematic drawing of the system considered in this paper. Here, FiI and M refer to a ferrimagnetic insulator and a heavy metal with local temperatures $T_{\rm F}$ and $T_{\rm M}$, respectively. The FiI layer consists of $A$ sublattice and $B$ sublattice. For sublattice spins $\bmS^A$ and $\bmS^B$, see the definition below Eq.~(\ref{eq:ndef01}). 
  }
  \label{fig:schematic}
\end{figure}

We consider a bilayer composed of a ferrimagnetic insulator (FiI) and a heavy metal (M) as shown in Fig.~\ref{fig:schematic}, where the FiI and M layers respectively have local temperatures $T_{\rm F}$ and $T_{\rm M}$. The FiI layer consists of two sublattices $A$ and $B$. Below, we introduce magnetization vector, 
\begin{equation}
  \bmm = \bmS^A + \bmS^B,
  \label{eq:mdef01}
\end{equation}
and the staggered magnetization (N\'{e}el) vector, 
\begin{equation}
  \bmn = \bmS^A - \bmS^B, 
  \label{eq:ndef01}
\end{equation}
where $\bmS^A =  {\cal N}^{-1} \sum_{i \in A} \bmS_i $, $\bmS^B =  {\cal N}^{-1} \sum_{i \in B} \bmS_i $, and ${\cal N}$ is the number of lattice sites in the system. For YIG, sublattice $A$ refers to the tetrahedral $d$ sites, while sublattice $B$ refers to the octahedral $a$ sites, respectively~\cite{Coey-text}. 

We consider the free energy for the FiI layer~\cite{Landau-electrodyn,Yamamoto19,Yamamoto22}, 
\begin{eqnarray}
  F_{\rm F} &=& \epsilon _0 v_0 \biggl[ \frac{u_2}{2}\bm{n}^2 + \frac{u_4}{4}(\bm{n}^2)^2
  + \frac{K_0}{2}(\bm{m}\times {\bf \hat{z}} )^2 
  \nonumber\\
  &+& \frac{r_0}{2} \bm{m}^2
  - L_{AB} \bmm \cdot \bmn \biggl]  
  - \gamma \hbar \bmH_{0} \cdot \bmm , 
\end{eqnarray}
where $u_2=(T-T^{(0)}_{\rm C})/T^{(0)}_{\rm C}$ is the quadratic coefficient with the bare transition temperature $T^{(0)}_{\rm C}$, $u_4$ is the quartic coefficient, $K_0$ describes the small uniaxial anisotropy, and $r_0$ gives the inverse of paramagnetic susceptibility. Besides, $\epsilon _0$ is the magnetic energy density, $v_0$ is the effective cell volume, $\mathfrak{h}_0 = \epsilon _0 v_0/(\gamma \hbar)$ is the unit of magnetic field with the gyromagnetic ratio $\gamma$ and Planck constant $\hbar$, and $\bmH_0= H_0 {\bf \hat{z}}$ is a static magnetic field. In this model, the phase transition is triggered by the N\'{e}el vector with the bare Curie temperature $T^{(0)}_{\rm C}$, and a smaller magnetization vector is induced by the coupling $L_{AB}$. Note that $\bmn$ is odd under the sublattice interchange operation $A \leftrightarrow B$ while $\bmm$ is even, such that the coefficient $L_{AB}$ is antisymmetric under the sublattice interchange operation, i.e., $L_{BA}= -L_{AB}$. 

We consider the free energy for the M layer~\cite{Yamamoto19,Yamamoto22} as 
\begin{equation}
  F_{\rm M} = \frac{1}{2 \chi_{\rm M}} \bmsig^2
  - \gamma \hbar \bmH_{0}  \cdot \bmsig, 
\end{equation}
where $\bmsig$ and $\chi_{\rm M}$ are respectively the spin accumulation and the static spin susceptibility of the M layer. We also consider the interaction at the FiI/M interface,
\begin{equation}
  F_{\rm F/M} = -J_m \bmm \cdot \bmsig - J_n \bmn \cdot \bmsig, 
\end{equation}
where $J_m$ ($J_n$) is hereafter referred to as the magnetic (N\'{e}el) coupling constant in this paper.

The SSE is driven by thermal fluctuations of $\bmm$, $\bmn$, and $\bmsig$, which are described by their dynamics. The dynamics of $\bmm$ and $\bmn$ in the FiI layer is described by the TDGL equations: 
\begin{eqnarray}
  \label{eq:TDGL_m01}
  \frac{\partial}{\partial t} \bm{m}
  &=& \gamma \bm{H}_m \times \bm{m}+\gamma \bm{H}_n \times \bm{n}+\Gamma _m \frac{\bm{H}_m}{\mathfrak{h}_0 } + {\bm \xi}, 
\\
  \label{eq:TDGL_n01}
    \frac{\partial}{\partial t} {\bm n}
  &=& \gamma {\bm H}_n \times {\bm m}+\gamma {\bm H}_m \times {\bm n}+\Gamma _n \frac{{\bm H}_n}{\mathfrak{h}_0} + {\bm \eta},
\end{eqnarray}
where $\Gamma _m$ and $\Gamma_n$ are damping coefficients for $\bmm$ and $\bmn$, respectively, and the effective fields $\bm{H}_m$ and $\bm{H}_n$ are defined by 
\begin{eqnarray}
  \bmH_m &=& -\frac{1}{\gamma \hbar} \frac{\partial}{\partial \bm{m}}
  \left( F_{\rm F}+F_{\rm F/M} \right), \\
  \bmH_n &=& -\frac{1}{\gamma \hbar} \frac{\partial}{\partial \bmn}
  \left( F_{\rm F} +F_{\rm F/M} \right).
\end{eqnarray}
In the above TDGL equations, the noise field ${\bm \xi}$ and ${\bm \eta}$ have zero mean, and due to the fluctuation-dissipation theorem they have variance 
\begin{eqnarray}
  \bra \xi^i (t) \xi^j (t') \ket &=& 
  \frac{2 \kB T_{\rm F} \Gamma_m}{\epsilon_0 v_0} \delta_{i,j} \delta(t-t'), \\
    \bra \eta^i (t) \eta^j (t') \ket &=& 
  \frac{2 \kB T_{\rm F} \Gamma_n}{\epsilon_0 v_0} \delta_{i,j} \delta(t-t'), 
\end{eqnarray}
where $\bra \cdots \ket$ represents averaging over noise field, $i$ and $j$ denote $x,y,z$, and there is no cross correlation between these two fields,
\begin{equation}
  \bra \xi^i (t) \eta^j (t') \ket = 0. 
\end{equation}
In the M layer, the dynamics of $\bmsig$ is described by the Bloch equation: 
\begin{equation}
  \frac{\partial}{\partial t} \bmsig
  = \gamma \bmH_{\sigma} \times \bmsig
  + \frac{\chi_{\rm M} \gamma \hbar}{\tau_{\rm M}} \bmH_{\sigma} + {\bm \zeta},
  \label{eq:Bloch01}
\end{equation}
where $\tau_{\rm M}$ is the spin-flip relaxation time of $\bmsig$, $\bmH_{\sigma}$ is given by 
\begin{equation}
  \bmH_{\sigma} = -\frac{1}{\gamma \hbar} \frac{\partial}{\partial \bmsig}
  \left( F_{\rm M} + F_{\rm F/M} \right), 
\end{equation}
and the noise field ${\bm \zeta}$ has zero mean and variance,
\begin{equation}
  \bra \zeta^i (t) \zeta^j (t') \ket = 
  \frac{2 \kB T_{\rm M} \chi_{\rm M} }{\tau_{\rm M}} \delta_{i,j} \delta(t-t'). 
\end{equation}

\subsection{Equilibrium properties}

Now, we discuss thermal equilibrium of the FiI and M layers. In the FiI layer, equilibrium values of $\bmm$ and $\bmn$ are determined by the conditions $\bmH_m= \bmH_n= {\bm 0}$. By setting 
\begin{eqnarray}
  \bmm_{\rm eq} &=& m_{\rm eq} {\bf \hat{z}}, \\
  \bmn_{\rm eq} &=& n_{\rm eq} {\bf \hat{z}}, 
\end{eqnarray}
we obtain 
\begin{eqnarray}
  m_{\rm eq} &=& \frac{L_{AB}}{r_0} n_{\rm eq}, \\
  n_{\rm eq} &=& \begin{cases} 
   \sqrt{\frac{1}{u_4} \left(1 +\frac{L_{AB}^2}{r_0} \right) 
     \frac{T_{\rm C} -T}{T_{\rm C}}} & T \le T_{\rm C}, \\
    0 & T > T_{\rm C}, 
    \end{cases}
\end{eqnarray}
in the limit $H_0 = 0$, where $T_{\rm C}$ is defined by 
\begin{equation}
  T_{\rm C} = T^{(0)}_{\rm C} \left( 1+ \frac{L_{AB}^2}{r_0} \right). 
\end{equation}
Under nonzero magnetic field $H_0 \neq 0$, since the analytical expression cannot be obtained, we numerically calculate $m_{\rm eq}$ and $n_{\rm eq}$ by using the conditions $\bmH_m= \bmH_n= {\bm 0}$.

\begin{figure}[t] 
  \begin{center}
    \includegraphics[width=8cm]{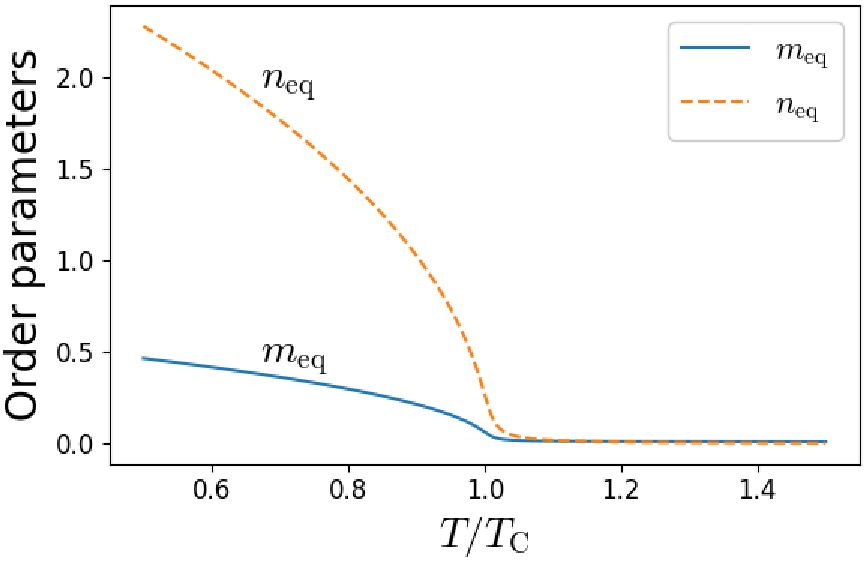}
  \end{center}
  \caption{
    Temperature dependence of $m_{\rm eq}$ and $n_{\rm eq}$
    for $H_0/\mathfrak{h}_0 = 0.01$, where $r_0 = 1.0$, $L_{AB}= 0.2$, and $u_4= 0.1$ are used.  }
  \label{fig:mneq01}
\end{figure}

Figure~\ref{fig:mneq01} shows equilibrium values of the order parameters $m_{\rm eq}$ and $n_{\rm eq}$ as a function of temperature for $H_0/\mathfrak{h}_0 = 0.01$, where $r_0 = 1.0$, $L_{AB}= 0.2$, and $u_4= 0.1$ are used. For YIG, because three spins $3 \bmS_d$ at the tetrahedral $d$ sites and two spins $2 \bmS_a$ at the octahedral $a$ sites align antiferromagnetically~\cite{Coey-text}, we have 
\begin{subequations} 
  \begin{align} 
    \bmS^A &= \frac{3}{5} \bmS_d, \\
    \bmS^B &= \frac{2}{5} \bmS_a = - \frac{2}{5} \bmS_d,
  \end{align}
\end{subequations}
where we used $\bmS_a = - \bmS_d$ in the equilibrium state. Therefore, from Eqs.~(\ref{eq:mdef01}) and (\ref{eq:ndef01}), we expect $m_{\rm eq}= (1/5){\bf \hat{z}} \cdot \langle \bmS_{d} \rangle $ and $n_{\rm eq}= {\bf \hat{z}} \cdot \langle \bmS_{d} \rangle$, from which we obtain $n_{\rm eq} = 5 m_{\rm eq}$. The parameters used to calculate Fig.~\ref{fig:mneq01} are chosen to reproduce this relationship.

Finally in the M layer, the equilibrium value of $\bmsig$ is determined by the condition $\bmH_{\sigma}= {\bm 0}$. By setting  
\begin{equation}
  \bmsig_{\rm eq} = \sigma_{\rm eq} {\bf \hat{z}}, 
\end{equation}
we obtain
\begin{eqnarray}
  \sigma_{\rm eq} = \chi_{\rm M} \Big( J_m m_{\rm eq} + J_n n_{\rm eq} \Big). 
\end{eqnarray}

\subsection{Nonequilibrium dynamics} 

Next, we investigate nonequilibrium fluctuations of $\bmm$ and $\bmn$ about $\bmm_{\rm eq}$ and $\bmn_{\rm eq}$. We substitute 
\begin{eqnarray}
  \bmm &=& \bmm_{\rm eq} + \delta \bmm, \label{eq:MeqdM01}  \\
  \bmn &=& \bmn_{\rm eq} + \delta \bmn, \label{eq:NeqdN01}
\end{eqnarray}
into the TDGL equations (\ref{eq:TDGL_m01}) and  (\ref{eq:TDGL_n01}), and linearize them with respect to $\delta \bmm$ and $\delta \bmn$. Then, we define the shorthand notation $\int_\omega = \int_{-\infty}^\infty \frac{d \omega}{2 \pi}$ and the Fourier transform $f(t)= \int_\omega f_\omega e^{-i \omega t}$, as well as we introduce the rotating coordinate representation 
\begin{equation}
  O^\pm = O^x \pm i O^y  
\end{equation}
for a vector variable ${\bm O}$. Now the minus branch of the linearized TDGL equation is transformed as~\cite{Yamamoto19,Yamamoto22} 
\begin{eqnarray}
(\omega-\widehat{\mathcal{A}}) 
\begin{pmatrix}
\delta m^{-}_\omega \\
\delta n^{-}_\omega \\
\end{pmatrix}
&=&
-\frac{1}{\hbar}
\begin{pmatrix}
{J_m} m_{\rm eq} + {J_n} n_{\rm eq} -  i \frac{J_m \Gamma_m }{\gamma \mathfrak{h}_0}\\
{J_n} m_{\rm eq} + {J_m} n_{\rm eq} -  i \frac{J_n \Gamma_n }{\gamma \mathfrak{h}_0}\\
\end{pmatrix} \delta \sigma^-_\omega \nonumber \\
&& +
\begin{pmatrix}
i \xi^-_{\omega} \\
i \eta^{-}_{\omega} \\
\end{pmatrix},
\label{eq:TDGLmatrix01}
\end{eqnarray}
where in the matrix 
\begin{equation}
  \widehat{\mathcal{A}} = 
\begin{pmatrix}
  a & b\\
  c & d \\
\end{pmatrix}, 
\end{equation}
each component is given by 
\begin{eqnarray}
  a &=& \gamma H_0 + \gamma \mathfrak{h}_0 K_0 m_{\rm eq} - i \Gamma_m (K_0+ r_0), \\
  b &=& i \Gamma_m L_{AB}, \\
  c &=& \gamma \mathfrak{h}_0 (K_0 + r) n_{\rm eq}+ i \Gamma_n L_{AB} , \\
  d &=& \gamma H_0 - \gamma \mathfrak{h}_0 r m_{\rm eq} - i \Gamma_n (r_0 - r),
\end{eqnarray}
where $r= r_0 - (u_2 + u_4 n^2_{\rm eq})$. Similarly, the plus branch of the TDGL equation is transformed as 
\begin{eqnarray}
(\omega + \widehat{\mathcal{A}}^*) 
\begin{pmatrix}
\delta m^{+}_\omega \\
\delta n^{+}_\omega \\
\end{pmatrix}
&=& \frac{1}{\hbar}
\begin{pmatrix}
{J_m} m_{\rm eq} + {J_n} n_{\rm eq} + i \frac{J_m \Gamma_m}{\gamma \mathfrak{h}_0}\\
{J_n} m_{\rm eq} + {J_m} n_{\rm eq} +  i \frac{J_n \Gamma_n}{\gamma \mathfrak{h}_0}\\
\end{pmatrix} \delta \sigma^{+}_\omega \nonumber \\
&& +
\begin{pmatrix}
i \xi^{+}_{\omega} \\
i \eta^{+}_{\omega} \\
\end{pmatrix}.
\label{eq:TDGLmatrix02}
\end{eqnarray}

Also, to investigate nonequilibrium fluctuations of $\bmsig$ about $\bmsig_{\rm eq}$, we substitute 
\begin{equation}
  \bmsig = \bmsig_{\rm eq} + \delta \bmsig \label{eq:SeqdS01}  
\end{equation}
into Eq.~(\ref{eq:Bloch01}) and linearize it with respect to $\delta \bmsig$. Then, moving into the rotating coordinate representation, the Bloch equation for $\delta \sigma_\omega^\pm$ is transformed as 
\begin{equation}
  \left( \omega+ \frac{i}{\tau_{\rm M}} \right) \delta \sigma^\pm_\omega
  =
  i \frac{\chi_{\rm M}}{\tau_{\rm M}} \left( {J_m } \delta m^\pm_\omega 
  +   {J_n} \delta n^\pm_\omega \right)
  + i \zeta^\pm_\omega.
\label{eq:Blochmatrix01}  
\end{equation}

\begin{figure}[t] 
  \begin{center}
    \includegraphics[width=8cm]{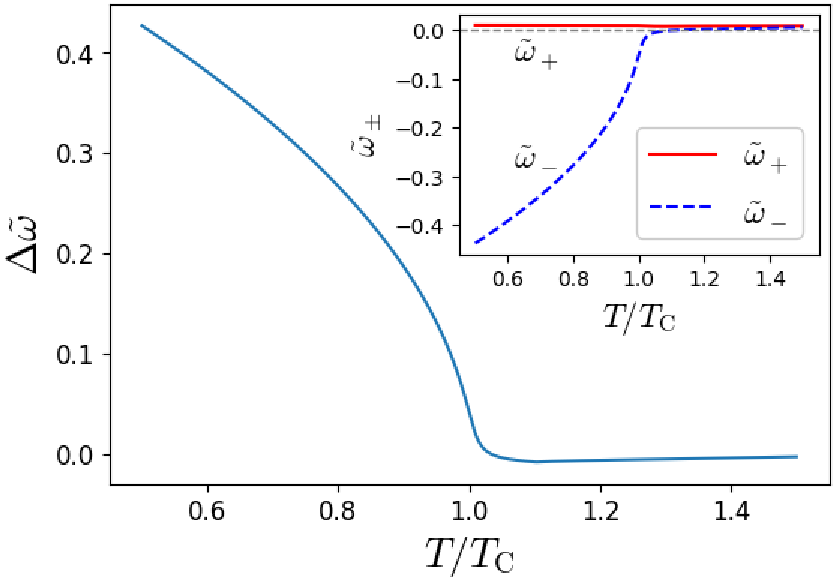}
  \end{center}
  \caption{
    Temperature dependence of the gap $\Delta \widetilde{\omega}= |\widetilde{\omega}_-| - |\widetilde{\omega}_+|$ between the acoustic mode $\omega_+$ and optical mode $\omega_-$, where the frequencies are normalized as $\widetilde{\omega}_\pm= \omega_\pm/(\gamma \mathfrak{h}_0)$. Here, parameters are chosen the same as in Fig.~\ref{fig:mneq01}, i.e., $H_0/\mathfrak{h}_0 = 0.01$, $r_0 = 1.0$, $L_{AB}= 0.2$, $u_4= 0.1$. Besides, we use $K_0 = 0.001$, $\Gamma_m/(\gamma \mathfrak{h}_0)= 0.01$, and $\Gamma_n/(\gamma \mathfrak{h}_0)= 0.01$. Inset: Temperature dependence of $\widetilde{\omega}_\pm$. 
  }
  \label{fig:Domega01}
\end{figure}

We first discuss the spin wave frequencies in our model. For this purpose, we consider the propagator $\widehat{\mathcal{G}} (\omega)= (\omega - \widehat{\cal A})^{-1}$, whose expression is given by~\cite{Yamamoto19,Yamamoto22} 
\begin{eqnarray}
  \widehat{\mathcal{G}} (\omega) &=& 
  \begin{pmatrix}
    \mathcal{G}_1 (\omega) & \mathcal{G}_2 (\omega) \\
    \mathcal{G}_3 (\omega) & \mathcal{G}_4 (\omega) \\
  \end{pmatrix} \nonumber \\
  &=&
  \frac{1}{(\omega- \lambda_+)(\omega- \lambda_-) }
  \begin{pmatrix}
    \omega- d  & b\\
    c & \omega - a \\
  \end{pmatrix},
  \label{eq:propagator01}
\end{eqnarray}
where 
\begin{equation}
  \lambda_\pm = \frac{a+ d \pm \sqrt{(a-d)^2+ 4 bc}}{2}. 
\end{equation}
The eigen-frequencies $\omega_\pm$ of the two modes are then given by
\begin{equation}
  \omega_\pm = {\rm Re} \; \lambda_\pm. 
\end{equation}
Note that, unlike Ref.~\cite{Yamamoto19}, we define $\omega_{\pm}$ so as to include the sign that represents the helicity of each magnon mode. Note also that $\omega_+ >0 $ has a character of acoustic (ferromagnetic) mode with {\it positive} helicity, whereas $\omega_- <0$ has a character of optical (antiferromagnetic) mode with {\it negative} helicity.

The inset of Fig.~\ref{fig:Domega01} shows temperature dependence of these two modes ${\omega}_\pm$ for $K_0= 0.001$, $\Gamma_m/(\gamma \mathfrak{h}_0)= 0.01$, and $\Gamma_n/(\gamma \mathfrak{h}_0)= 0.01 $, where other parameters are the same as in Fig.~\ref{fig:mneq01}, and the frequencies are normalized as $\widetilde{\omega}_\pm= \omega_\pm/(\gamma \mathfrak{h}_0)$. From the figure we see that below $T_{\rm C}$ the optical mode $|\omega_-|$ develops roughly proportional to $n_{\rm eq}$, while the acoustic mode $|\omega_+|$ remains almost constant. Also, in Fig.~\ref{fig:Domega01}, we plot $\Delta \widetilde{\omega} = |\widetilde{\omega}_-| - |\widetilde{\omega}_+ |$ as a function of temperature. Note that this figure should be compared with Fig.~2 of Ref.~\cite{Barker16}. 

\subsection{Spin current injected into heavy metal}

Now we are in a position to calculate the spin current injected into the M layer. This quantity is defined by~\cite{Adachi11,Adachi13} 
\begin{equation}
  I_{\rm s}(t) = \frac{\partial}{\partial t} \bra \sigma^z(t) \ket \Big]_{\rm interface},
    \label{eq:js01}
\end{equation}
where the symbol $\left. \cdots \right]_{\rm interface}$ means to specify the rate of change due to the interfacial spin transfer.
  Then, using the $z$-component of the Bloch equation (\ref{eq:Bloch01}), the injected spin current is transformed as~\cite{Adachi18,Yamamoto19}
\begin{equation}
  I_{\rm s} = \frac{1}{\hbar}
  \int_\omega 
  {\rm Im} \Big( 
  {J_m}  \bra \bra \delta m^-_\omega \delta \sigma^+_{-\omega} \ket \ket
  +
  {J_n}  \bra \bra \delta n^-_\omega \delta \sigma^+_{-\omega} \ket \ket
  \Big), 
\end{equation}
where the quantity $\bra \bra \cdots \ket \ket$ is defined by $\bra \delta m^-_\omega \delta \sigma^+_{\omega'} \ket = 2 \pi \delta(\omega+\omega') \bra \bra \delta m^-_\omega \delta \sigma^+_{-\omega} \ket \ket$. The above equation means that the injected spin current can be obtained by evaluating the interface correlations $\bra \bra \delta m^-_\omega \delta \sigma^+_{-\omega} \ket \ket$ and  $\bra \bra \delta n^-_\omega \delta \sigma^+_{-\omega} \ket \ket$.

In order to calculate these two interface correlations, we solve the TDGL equations (\ref{eq:TDGLmatrix01}) and (\ref{eq:TDGLmatrix02}) and Bloch equation (\ref{eq:Blochmatrix01}) perturbatively with respect to $J_m$ and $J_n$, and evaluate $\bra \bra \delta m^-_\omega \delta \sigma^+_{-\omega} \ket \ket$ and  $\bra \bra \delta n^-_\omega \delta \sigma^+_{-\omega} \ket \ket$. Using $\bra \bra \xi^-_\omega \xi^+_{-\omega} \ket \ket = 4 \kB T_{\rm F} \Gamma_m/(\epsilon_0 v_0)$, $\bra \bra \eta^-_\omega \eta^+_{-\omega} \ket \ket = 4 \kB T_{\rm F} \Gamma_n/(\epsilon_0 v_0)$, and $\bra \bra \zeta^-_\omega \zeta^+_{-\omega} \ket \ket = 4 \kB T_{\rm M} \chi_{\rm M}/\tau_{\rm M}$, and after summarizing the result up to the second order with respect to $J_m$ and $J_n$, the injected spin current is expressed as 
\begin{eqnarray}
  I_{\rm s} &=&
  \frac{4 \chi_{\rm M} \kB \Delta T}{\epsilon_0 v_0 \hbar \tau_{\rm M}}
  \Big( J_m^2 {\cal L}_m + J_n^2 {\cal L}_n + 2 J_m J_n {\cal L}_{mn} \Big),
  \label{eq:Is_final01}
\end{eqnarray}
where $\Delta T= T_{\rm M}- T_{\rm F}$. In the above equation, ${\cal L}_m$, ${\cal L}_n$, and ${\cal L}_{mn}$ are defined by
\begin{eqnarray}
  {\cal L}_m &=&
  \int_\omega \Big( \Gamma_m |{\cal G}_1(\omega)|^2
  + \Gamma_n |{\cal G}_2 (\omega)|^2\Big) \omega |g(\omega)|^2, \\
  {\cal L}_n &=&
  \int_\omega \Big(\Gamma_n |{\cal G}_4 (\omega)|^2
  +  \Gamma_m |{\cal G}_3 (\omega)|^2 \Big) \omega |g(\omega)|^2, \\
  {\cal L}_{mn} &=&
  \int_\omega {\rm Re} \Big( \Gamma_m {\cal G}_1 (\omega) \left[ {\cal G}_3(\omega) \right]^*  
  + \Gamma_n {\cal G}_4(\omega) \left[ {\cal G}_2(\omega) \right]^* 
  \Big) \nonumber \\
  && \qquad \times \omega |g(\omega)|^2, 
\end{eqnarray}
where $g(\omega)= (\omega + i/\tau_{\rm M})^{-1}$.

Note that, in the limit $\omega_\pm \tau_{\rm M} \ll 1$, the above result can be expressed by using the spin-mixing conductance for ferrimagnets~\cite{Kamra17}. Indeed, using the spectral representation $ {\bf \hat{z}} \cdot \bra \bmm \times \partial_t \bmm \ket = \int_\omega \omega \bra \bra \delta m_\omega^- \delta m_{-\omega}^+ \ket \ket$, ${\bf \hat{z}} \cdot \bra \bmn \times \partial_t \bmn \ket= \int_\omega \omega \bra \bra \delta n_\omega^- \delta n_{-\omega}^+ \ket \ket$, and $ {\bf \hat{z}} \cdot \bra \bmm \times \partial_t \bmn + \bmn \times \partial_t \bmm \ket= \int_\omega \omega \bra \bra \delta m_\omega^- \delta n_{-\omega}^+ + \delta n_\omega^- \delta m_{-\omega}^+ \ket \ket$, the above result can be expressed as
  \begin{equation}
    I_s = I_s^{\rm pump} (T_{\rm F}) - I_s^{\rm back} (T_{\rm M}), 
  \end{equation}
  where the pumping current at temperature $T$ is given by   
\begin{eqnarray}
  I_{\rm s}^{\rm pump} (T) &=&
  -G_m {\bf \hat{z}} \cdot \bra \bmm \times \partial_t \bmm \ket 
  - 
  G_n {\bf \hat{z}} \cdot \bra \bmn \times \partial_t \bmn \ket \nonumber \\
  && -
  G_{mn} {\bf \hat{z}} \cdot \bra \bmm \times \partial_t \bmn + \bmn \times \partial_t \bmm \ket, 
  \label{eq:Gs01}
\end{eqnarray}
and the backflow is given by $I_s^{\rm back} (T_{\rm M})= I_s^{\rm pump} (T=T_{\rm M})$ due to the fluctuation-dissipation theorem~\cite{Foros05}. In Eq.~(\ref{eq:Gs01}), the spin-mixing conductance is given by $G_m = J_m^2 \chi_{\rm M} \tau_{\rm M}/\hbar$, $G_n = J_n^2 \chi_{\rm M} \tau_{\rm M}/\hbar$, and $G_{mn} = J_m J_n \chi_{\rm M} \tau_{\rm M}/\hbar$, where the minus sign in each term represents the fact that the mangon carries spin $-\hbar$ (see Eq.~(27) in \cite{Adachi13}).

Below, we calculate the above integral over $\omega$ by picking up the magnon poles $\omega= \lambda_\pm^*$. Then, ${\cal L}_m$ is transformed as 
\begin{eqnarray}
  {\cal L}_m &=&
  {\cal L}_m^{(+)}+    {\cal L}_m^{(-)}, \label{eq:Lm_pm01} \\
  {\cal L}_m^{(+)} &=& {\rm Re} \left( \frac{i \lambda_+^*}{D_+(\lambda_+^*)} f_m (\lambda_+^*) \right), \label{eq:Lm_p01}\\
  {\cal L}_m^{(-)} &=& {\rm Re} \left( \frac{i \lambda_-^*}{D_- (\lambda_-^*)} f_m (\lambda_-^*) \right),  \label{eq:Lm_m01} 
\end{eqnarray}
where 
\begin{eqnarray}
  f_m (\lambda) &=& \Gamma_m(\lambda-d)(\lambda-d^*) + \Gamma_n |b|^2, \\
  D_+(\lambda) &=& (\lambda- \lambda_+)(\lambda- \lambda_-)(\lambda- \lambda^*_-) (\lambda^2+ \tau_{\rm M}^{-2}), \\
    D_-(\lambda) &=& (\lambda- \lambda_+)(\lambda- \lambda^*_+)(\lambda- \lambda_-) (\lambda^2+ \tau_{\rm M}^{-2}), \label{eq:Dm01}
\end{eqnarray}
and taking the real part in Eqs.~(\ref{eq:Lm_p01}) and (\ref{eq:Lm_m01}) is due to our approximation of picking up only magnon poles. Note that ${\cal L}_m^{(+)}$ represents the contribution from the acoustic (ferromagnetic) magnons with $\omega_+$, whereas ${\cal L}_m^{(-)}$ represents the contribution from the optical (antiferromagnetic) magnons with $\omega_-$. In a similar manner, ${\cal L}_n$ is transformed as
\begin{eqnarray}
  {\cal L}_n &=&
  {\cal L}_n^{(+)}+    {\cal L}_n^{(-)}, \\
  {\cal L}_n^{(+)} &=& {\rm Re} \left( \frac{i \lambda_+^*}{D_+(\lambda_+^*)} f_n (\lambda_+^*) \right), \\
  {\cal L}_n^{(-)} &=& {\rm Re} \left( \frac{i \lambda_-^*}{D_- (\lambda_-^*)} f_n (\lambda_-^*) \right) ,  
\end{eqnarray}
where
\begin{equation}
  f_n (\lambda) = \Gamma_n (\lambda-a)(\lambda-a^*) + \Gamma_m |c|^2, 
\end{equation}
and ${\cal L}_{mn}$ is transformed as 
\begin{eqnarray}
  {\cal L}_{mn} &=&
  {\cal L}_{mn}^{(+)}+    {\cal L}_{mn}^{(-)}, \\  
  {\cal L}_{mn}^{(+)} &=& {\rm Re} \left( \frac{i \lambda_+^*}{D_+(\lambda_+^*)} f_{mn} (\lambda_+^*) \right), \\
  {\cal L}_{mn}^{(-)} &=& {\rm Re} \left( \frac{i \lambda_-^*}{D_-(\lambda_-^*)} f_{mn} (\lambda_-^*) \right), 
\end{eqnarray}
where
\begin{equation}
  f_{mn} (\lambda) = \Gamma_m  (\lambda-d)c^* + \Gamma_n (\lambda-a)b^* . 
\end{equation}
Substituting these expressions into Eq.~(\ref{eq:Is_final01}), we can calculate the spin current driven by the SSE.

\begin{figure} [t] 
  \begin{center}
    \includegraphics[width=8.5cm]{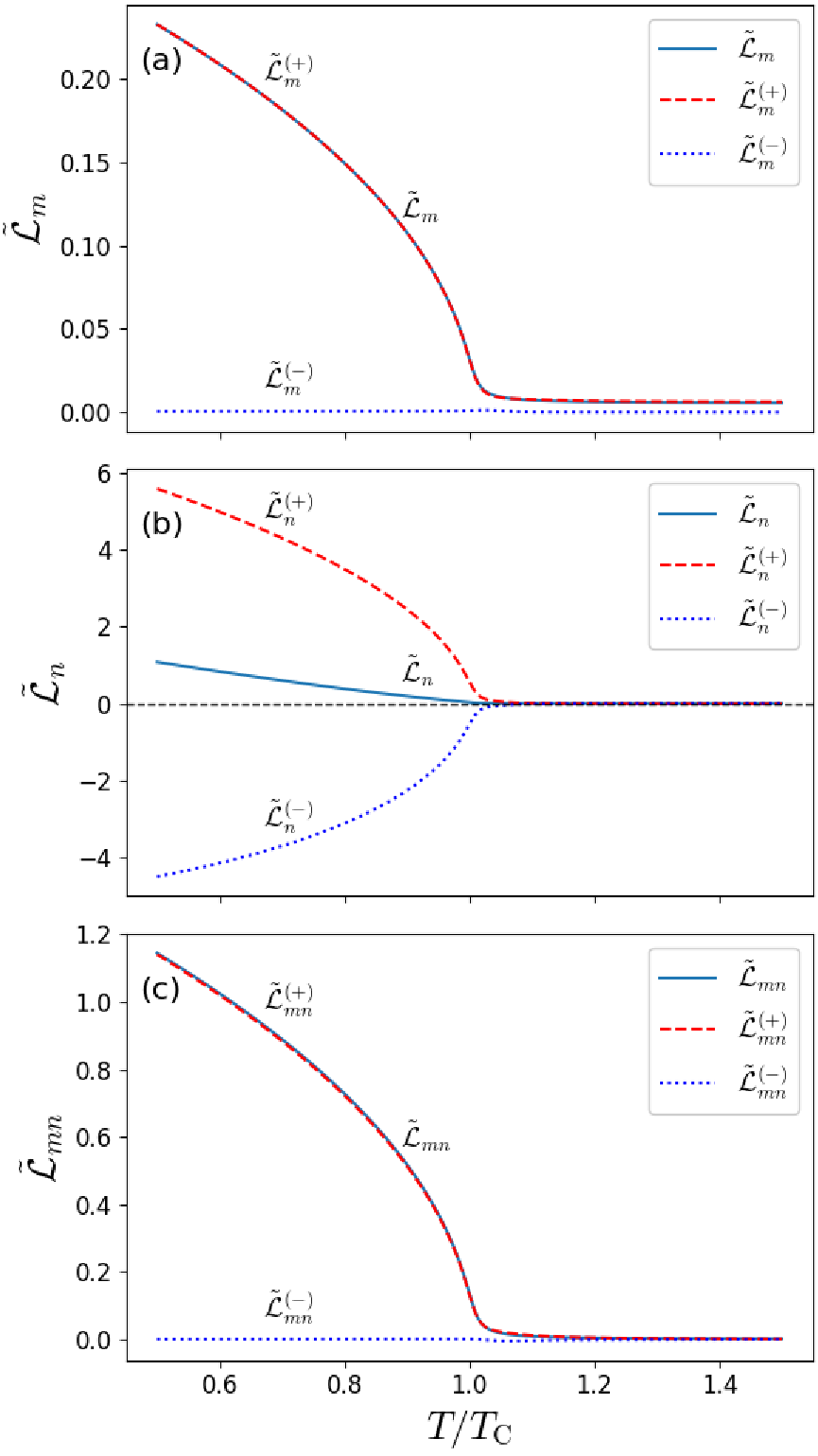}
  \end{center}
  \caption{
    Temperature dependence of (a) $\widetilde{\cal L}_m = {\cal L}_m \gamma \mathfrak{h}_0$ and $\widetilde{\cal L}_m^{(\pm)} = {\cal L}_m^{(\pm)} \gamma \mathfrak{h}_0$,
    (b) $\widetilde{\cal L}_n = {\cal L}_n \gamma \mathfrak{h}_0$ and $\widetilde{\cal L}_n^{(\pm)} = {\cal L}_n^{(\pm)} \gamma \mathfrak{h}_0 $, and 
    (c) $\widetilde{\cal L}_{mn} = {\cal L}_{mn} \gamma \mathfrak{h}_0 $ and $\widetilde{\cal L}_{mn}^{(\pm)} = {\cal L}_{mn}^{(\pm)} \gamma \mathfrak{h}_0 $.
    Here, the same parameters as Fig.~\ref{fig:Domega01} are used, and we set $\gamma \mathfrak{h}_0 \tau_{\rm M} = 1.0$. 
  }
  \label{fig:L_mnmn01}
  \end{figure}

\section{Spin Seebeck effect near $T_{\rm C}$ \label{sec:III}}

\begin{figure}[t] 
  \begin{center}
        \includegraphics[width=8.5cm]{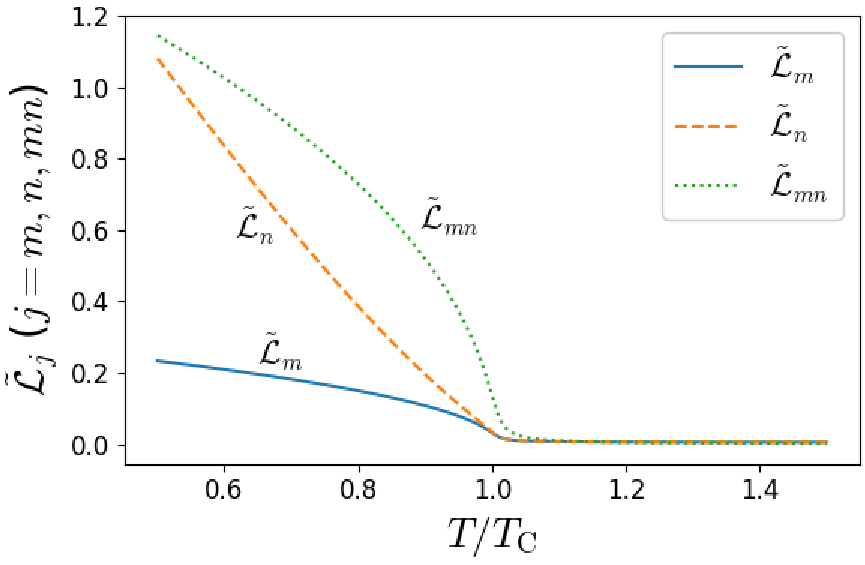}
  \end{center}
  \caption{ Temperature dependence of $\widetilde{\cal L}_m= {\cal L}_m \gamma \mathfrak{h}_0$, $\widetilde{\cal L}_n = {\cal L}_n \gamma \mathfrak{h}_0$, and $\widetilde{\cal L}_{mn} = {\cal L}_{mn} \gamma \mathfrak{h}_0$ in Fig.~\ref{fig:L_mnmn01} are shown on the same scale. 
  }
    \label{fig:L01}
\end{figure}

Now, we are ready to discuss the SSE signal near $T_{\rm C}$. This quantity, expressed by Eq.~(\ref{eq:Is_final01}), is divided into three contributions: ${\cal L}_m$, ${\cal L}_n$, and ${\cal L}_{mn}$. 

We first examine temperature dependence of ${\cal L}_m$. Figure \ref{fig:L_mnmn01}(a) shows temperature dependence of $\widetilde{\cal L}_m = {\cal L}_m \gamma \mathfrak{h}_0$ for the same parameters used in Fig.~\ref{fig:Domega01}, where we additionally set $\gamma \mathfrak{h}_0 \tau_{\rm M}= 1.0$. This quantity can be decomposed into two parts as shown in Eq.~(\ref{eq:Lm_pm01}), where ${\cal L}_m^{(+)}$ represents a contribution from the acoustic (ferromagnetic) magnons, whereas ${\cal L}_m^{(-)}$ is a contribution from the optical (antiferromagnetic) magnons. From the figure we see that ${\cal L}_m$ is essentially determined by ${\cal L}_m^{(+)}$, which means that the acoustic (ferromagnetic) magnons play decisive role in ${\cal L}_m$. In this case, the temperature dependence near $T_{\rm C}$ has a convex upward curvature, whose behavior is similar to the equilibrium magnetization $m_{\rm eq}$. This is consistent with the previous results taking account only of the acoustic (ferromagnetic) magnons through the magnetic coupling $J_m$~\cite{Barker16,Adachi18}.

We next examine temperature dependence of ${\cal L}_n$. Figure \ref{fig:L_mnmn01}(b) shows temperature dependence of $\widetilde{\cal L}_n = {\cal L}_n \gamma \mathfrak{h}_0$. As in ${\cal L}_m$, the plus branch ${\cal L}_n^{(+)}$ represents a contribution from the acoustic (ferromagnetic) magnons, whereas the minus branch ${\cal L}_n^{(-)}$ is a contribution from the optical (antiferromagnetic) magnons. From the figure we see that ${\cal L}_n^{(+)}$ is positive with convex upward curvature whereas ${\cal L}_n^{(-)}$ is negative with convex downward curvature, giving in total a positive contribution to ${\cal L}_n$ with slightly convex downward curvature. We also find that the magnitude of ${\cal L}_n$ is approximately five times as large as ${\cal L}_m$. 

Finally, we examine temperature dependence of ${\cal L}_{mn}$. Figure \ref{fig:L_mnmn01}(c) shows temperature dependence of $\widetilde{\cal L}_{mn} = {\cal L}_{mn} \gamma \mathfrak{h}_0$. As seen from the figure, ${\cal L}^{(-)}_{mn}$ is negligibly small in comparison to ${\cal L}^{(+)}_{mn}$, and ${\cal L}^{(+)}_{mn}$ is dominating the behavior of ${\cal L}_{mn}$. Also we find that overall temperature dependence of ${\cal L}_{mn}$ is quite similar to ${\cal L}_{m}$, but the magnitude of ${\cal L}_{mn}$ is roughly five times larger than ${\cal L}_m$. 

In Fig.~\ref{fig:L01}, temperature dependencies of ${\cal L}_m$, ${\cal L}_n$, and ${\cal L}_{mn}$ are shown in the same figure. From Eq.~(\ref{eq:Is_final01}), we notice that in an extreme case in the absence of N\'{e}el coupling (${J}_n =0$), temperature dependence of $I_{\rm s}$ is determined by ${\cal L}_m$ alone. Then, the result becomes similar to the case in Ref.~\cite{Adachi18} where the SSE in a simple ferromagnet is discussed by the GL theory. Likewise, in another extreme case in the absence of magnetic coupling (${J}_m =0$), temperature dependence of $I_{\rm s}$ is determined by ${\cal L}_n$ alone. Although the temperature dependence in this latter case becomes slightly convex downward, the curvature is too small to explain the experiment~\cite{Uchida14b} (see Fig.~3 therein).

Now, having the SSE experiment for a YIG/Pt system in mind~\cite{Uchida14b}, we examine temperature dependence of the SSE signal $I_{\rm s}$. In doing so, it is convenient to rewrite Eq.~(\ref{eq:Is_final01}) as
\begin{equation}
  I_{\rm s} =
  K J_m^2 \left[  {\cal L}_m + \left( \frac{J_n}{J_m} \right)^2 {\cal L}_n
  + 2 \left( \frac{J_n}{J_m} \right) {\cal L}_{mn} \right],
  \label{eq:Is_final02}
\end{equation}
where $K= {4 \chi_{\rm M} \kB \Delta T}/({\epsilon_0 v_0 \hbar \tau_{\rm M}})$. Then, if we normalize the spin current by its value at $T= 0.5T_{\rm C}$, we find that temperature dependence of the SSE signal is determined only by the ratio $J_n/J_m$. Below, we examine temperature dependence of the SSE signal $I_{\rm s}(T)/I_{\rm s}(0.5T_{\rm C})$ by varying the value of $J_n/J_m$.

\begin{figure}[t] 
  \begin{center}
        \includegraphics[width=8.5cm]{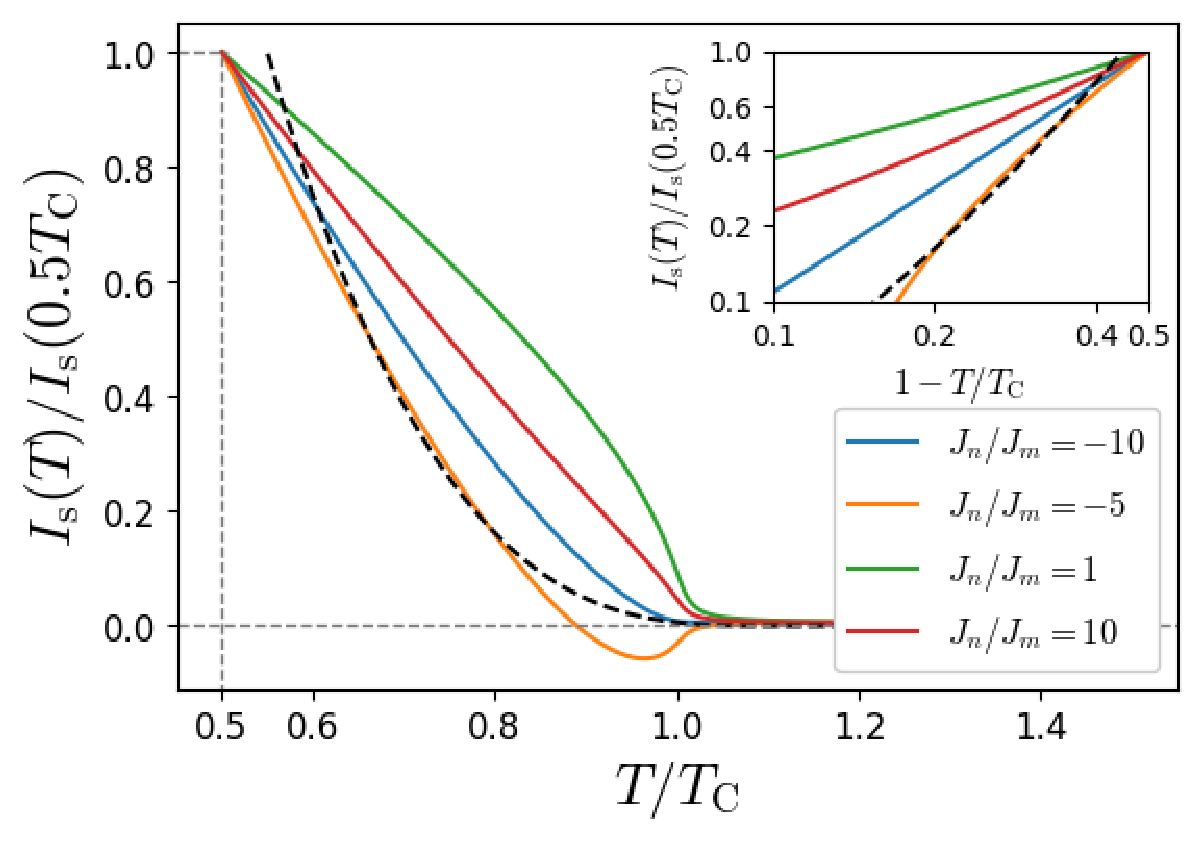}
  \end{center}
  \caption{
    Temperature dependence of the SSE signal $I_{\rm s}$ for several different values of $J_n/J_m$, where the data is normalized by its value at $T= 0.5 T_{\rm C}$. The black dashed curve represents a fitting to $J_n/J_m= -5.0$ data by $I_{\rm s} \propto (1-T/T^*_{\rm C})^3$ with $T^*_{\rm C}= 1.1T_{\rm C}$. Here, the same parameters as Fig.~\ref{fig:Domega01} are used, and we set $\gamma \mathfrak{h}_0 \tau_{\rm M} = 1.0$. Inset: Double logarithmic plot of $I_{\rm s}$. 
  }
  \label{fig:SSE01}
\end{figure}

Figure~\ref{fig:SSE01} shows our main result, where the SSE signal ${I}_{\rm s}$ is plotted as a function of temperature for several choices of $J_n/J_m$. When the strength of the N\'{e}el coupling is comparable to the magnetic coupling (green curve, $J_n/J_m = 1$), the SSE signal shows convex upward temperature dependence near $T_{\rm C}$ as seen in the result of Ref.~\cite{Barker16,Adachi18}. When the strength of the N\'{e}el coupling becomes much larger than the magnetic coupling (red curve, $J_n/J_m = 10$), the SSE signal changes into a $T$-linear dependence. Moreover, when the sign of the N\'{e}el coupling becomes opposite to the magnetic coupling while keeping the absolute value of the N\'{e}el coupling much larger than the magnetic coupling (blue curve, $J_n/J_m = -10$), the SSE signal shows convex downward temperature dependence. Finally, when the absolute value of the N\'{e}el coupling becomes slightly smaller than the last case while keeping $J_m J_n < 0$ (orange curve, $J_n/J_m = -5$), the SSE signal has a negative value just below $T_{\rm C}$ and shows more convex downward temperature dependence. Note that the SSE signal of Ref.~\cite{Uchida14b} appears significantly negative just below $T_{\rm C}$ (see Fig.~3 therein).  

In addition to the result shown in Fig.~\ref{fig:SSE01}, we vary the parameter $J_n/J_m$ and systematically examine temperature dependence of ${I}_{\rm s}$. Then, for $-5 \!\alt\! J_n/J_m \!\alt\! -0.1$, the temperature region of negative $I_{\rm s}(T)$ becomes too wide in comparison to the experiment. For $-0.1 \!\alt\!  J_n/J_m  \!\alt\! 10$, the temperature dependence becomes convex upward, and for $J_n/J_m \!\alt\! -10$ or $10  \!\alt\! J_n/J_m$, the $I_{\rm s}(T)$ curve exhibits almost $T$-linear dependence. Since the experimental data shows $I_{\rm s} \propto (1- T/T_{\rm C})^3$ power law~\cite{Uchida14b}, we try to fit the calculated SSE signal by this curve. Then, we find that introducing a shifted Curie temperature $T^*_{\rm C}= 1.1 T_{\rm C}$ improves the fitting, and the result is shown in Fig.~\ref{fig:SSE01} as the black dashed curve. From the inset of Fig.~\ref{fig:SSE01} as well as considering that a small negative value of the SSE is experimentally visible for $0.8 \!\alt\! T/T_{\rm C} \!\alt\! 1.0$, we conclude that the experimental data can be phenomenologically described by the parameter set $J_n/J_m = -5.0$.

\section{Discussion and Conclusion \label{sec:IV}} 

By comparing the SSE experiment for YIG/Pt sample in the vicinity of $T_{\rm C}$ (Fig.~3 in Ref.~\cite{Uchida14b}) with our theoretical result, we conclude that the two exchange coupling constants of $J_m$ and $J_n$ need to have the opposite signs and satisfy $J_n/J_m \approx -5.0$ in order to reproduce the experiment. Although the phenomenological GL approach to the ferrimagnetic SSE is our main focus and therefore a precise argument on the microscopic detail of such exchange coupling is beyond our scope, below we would like to discuss a possible origin for such condition. 

We first note that, applying the present GL model to YIG, the tetrahedral $d$ sites are assigned as $A$ sublattice and the octahedral $a$ sites are assigned as $B$ sublattice, respectively~\cite{Coey-text}. Then, the magnetization vector $\bmm$ and N\'{e}el vector $\bmn$ [Eqs.~(\ref{eq:mdef01}) and (\ref{eq:ndef01})] are represented as
\begin{subequations} 
  \begin{align} 
    \bmm &= \frac{1}{5} \left( 3 \bmS_d + 2 \bmS_a \right), \\
    \bmn &= \frac{1}{5} \left( 3 \bmS_d - 2 \bmS_a \right). 
  \end{align}
\end{subequations}
We solve the above equations for $\bmS_d$ and $\bmS_a$, and evaluate the interfacial exchange interaction. Then, for three $\bmS_d$ spins at tetrahedral $d$ sites and two $\bmS_a$ spins at octahedral $a$ sites, we obtain 
\begin{equation}
  J_d \bmsig \cdot 3 \bmS_d +   J_a \bmsig \cdot 2 \bmS_a
  =
  J_m \bmsig \cdot \bmm +   J_n \bmsig \cdot \bmn,
  \label{eq:Exchange01}
\end{equation}
where $J_d$ ($J_a$) is the exchange coupling constant at the $d$ site ($a$ site), and $J_m$ and $J_n$ are given by 
\begin{subequations} 
  \begin{align} 
    J_m &= \frac{5}{2} \Big( J_d + J_a \Big), \label{eq:Jm01}\\
    J_n &= \frac{5}{2} \Big( J_d - J_a \Big). \label{eq:Jn01}
  \end{align}
\end{subequations}
Equations~(\ref{eq:Jm01}) and (\ref{eq:Jn01}) mean that, if $J_d$ and $J_a$ have the {\it same sign}, we obtain $-1 < J_n/J_m < 1$. Conversely, if $J_d$ and $J_a$ have the {\it opposite sign}, the condition $J_n/J_m \approx -5.0$ that is consistent with experimental finding~\cite{Uchida14b} can be realized. For example, if $ J_d = -2 J_0 $ and $J_a = 3 J_0$ for a certain value $J_0$, we obtain $J_n/J_m = -5.0$.  

In order to discuss a possible microscopic origin of the situation considered above, we refer to the knowledge accumulated in the context of Kondo problem~\cite{Kondo-review} where the exchange coupling between a localized spin and conduction-electron spins ($s$-$d$ exchange coupling) has been discussed extensively. As is known in the literature~\cite{Coey-text,Nozieres80}, the sign and magnitude of the $s$-$d$ exchange coupling depend sensitively on the local environment (see Sec.~5.3 of Ref.~\cite{Coey-text}). For example if we limit ourselves to an Fe impurity, when diluted in Pd, the sign of the $s$-$d$ exchange coupling between Fe impurity spin and Pd conduction-electron spin is positive (ferromagnetic), leading to the formation of a giant moment~\cite{Clogston62}. By contrast when diluted in Mo, the sign of the $s$-$d$ exchange coupling between Fe impurity spin and Mo conduction-electron spin is negative (antiferromagnetic), leading to the Kondo effect~\cite{Sarachik64}. In the present case of Fe spins at the YIG/Pt interface, because the crystal fields between the tetrahedral $d$ sites and octahedral $a$ sites are different, the local electronic environments are also different. Therefore we argue that for the exchange coupling $J_d$ ($J_a)$ between conduction-electron spin in Pt and Fe spin at tetrahedral $d$ sites (octahedral $a$ sites), we can have a situation where $J_d$ and $J_a$ have opposite signs, giving rise to the condition $J_n/J_m \approx -5.0$.

  Before ending, we would like to comment on the influence of interface roughness on the N\'{e}el coupling $J_n$ in Eq.~(\ref{eq:Exchange01}). Since an interface of a real sample used for experiments is exposed to roughness, the roughness influences the interfacial exchange interaction. Then, Eq.~(\ref{eq:Exchange01}) should be taken to hold on average, and the $J_n$ value should be understood as phenomenological one that is defined after taking account of the random average. With this understanding, we think the influence of interface roughness on the $J_n$ value in Eq.~(\ref{eq:Exchange01}) is relatively weak. The reasoning is as follows. As discussed above, the sign of $J_d$ and $J_a$ that controls the magnitude of $J_n$ at YIG/Pt interface is determined by the local electronic environments of Fe spin such as the crystal fields at each site, where the tetrahedral $d$ site (sublattice $A$) and the octahedral $a$ site (sublattice $B$) are not equivalent. Therefore, the compensated and uncompensated aspects of YIG/Pt interface does not seem to play a decisive role in determining $J_n$, and hence the $J_n$ value is expected to be affected rather weakly by the roughness of YIG/Pt interface. This is in stark contrast to the antiferromagnetic SSE, where the sublattices $A$ and $B$ are equivalent and thus a spin-uncompensated interface is necessary to obtain a sizable $J_n$. That is, the N\'{e}el coupling $J_n$ in the latter case of antiferromagnetic SSE is highly susceptible to the interface roughness. Indeed, the strange sign change of the antiferromagnetic SSE observed in Cr$_2$O$_3$/Pt across the spin-flop transition~\cite{Li20}, which we explained in terms of the sizable N\'{e}el coupling~\cite{Yamamoto22}, vanishes after the Cr$_2$O$_3$/Pt interface is etched (see Extended Data Fig.~7 of \cite{Li20}). In passing, we note that the spin-flip process of conduction electrons proposed in Ref.~\cite{Masuda24} applies to the SSE under spin canted configuration and hence has no influence on the present problem.

To summarize this paper, GL theory of the SSE in two-sublattice ferrimagnets near $T_{\rm C}$ has been developed. We have pointed out the importance of the N\'{e}el coupling, i.e., the interfacial exchange coupling between the N\'{e}el vector in the ferrimagnet and spin accumulation in the heavy metal in understanding the convex downward temperature dependence of the SSE observed in YIG/Pt system near $T_{\rm C}$~\cite{Uchida14b}. The N\'{e}el coupling was proposed in the early stage of antiferromagnetic spintronics~\cite{Takei15} [see Eq. (1) therein], and the same coupling was used to explain the strange sign change of the antiferromagnetic SSE across the spin-flop transition~\cite{Yamamoto22}. This work has demonstrated the importance of the N\'{e}el coupling in developing ferrimagnetic spintronics~\cite{Kim22}.

\acknowledgments 
We are grateful to J. Otsuki for discussion on the $s$-$d$ exchange interaction, Y. Yamamoto and M. Teranishi for discussion in the early stage of the work. This work was financially suported by JSPS KAKENHI Garnts No. JP23K23209 and No. JP23K21077.




\end{document}